\documentstyle[prd,preprint,aps]{revtex}
\newcommand{\be}{\begin{equation}}
\newcommand{\ee}{\end{equation}}
\newcommand{\bea}{\begin{eqnarray}}
\newcommand{\eea}{\end{eqnarray}}

\newcommand{\V} {{\cal V}_n}
\newcommand{\Y} {{\overline \psi}}

\begin{document}

\reversemarginpar
\tighten

\title{Holography: 2-D or not 2-D?} 

\author {A.J.M. Medved}
%\thanks{E-mail:~joey_medved@mcs.vuw.ac.nz}

\address{
School of Mathematical and Computing Sciences\\
Victoria University of Wellington\\
PO Box 600, Wellington, New Zealand \\
E-Mail: joey\_medved@mcs.vuw.ac.nz}

\maketitle

\begin{abstract}

As was  recently pointed out  by Cadoni, 
 a certain class
of two-dimensional gravitational theories will exhibit  (black hole) 
thermodynamic
behavior that is reminiscent of a free field theory. In the
current letter,  a direct correspondence is established between 
these two-dimensional models 
 and the strongly curved regime of (arbitrary-dimensional)  
anti-de Sitter  gravity. On this basis, we go on to speculatively argue
that two-dimensional gravity may ultimately be utilized 
for identifying  and  perhaps even  understanding holographic
dualities.
\\
\end{abstract}

There is a growing suspicion that the holographic principle
may  be a crucial element in linking together semi-classical gravity
and   the fundamental quantum theory.  That is to say, the holographic
storage of information can, perhaps,  be viewed
as a semi-classical manifestation of some deep,
fundamental principle that has its origins in 
the (yet-to-be-understood) quantum nature of spacetime.
(For a review on the holographic principle, see \cite{bouxxx}.
For a discussion on how it might connect with quantum gravity,
see \cite{smoxxx}.)  

The essence of this holographic paradigm is that
the entropy (or, equivalently, the accessible information)
in a given region of spacetime should have a precise limit
which  can be formulated in
terms of the  ``area'' of a suitably defined surface.  
[For  a $d$-dimensional spacetime, this ``area'' would measure
the volume of 
some  $(d-2)$-dimensional hypersurface.]   In the ``conventional''
({\it i.e.}, flat-space) quantum
world, such 
a bound has   contradictory implications;
for instance,   quantum field theory predicts that the 
entropy will vary extensively with the volume of the applicable region.
Nonetheless,   such   expectations  need not persist once
gravitational interactions have been ``turned on''.
Indeed, the most strongly gravitating of objects --- black
holes ---  have a clear thermodynamic interpretation \cite{Bek,Haw} which
necessarily implies that $S_{max}\propto A$ \cite{tho}; 
with $S_{max}$ being  the maximal
amount of entropy that can be stored in a region bounded
by a surface of area $A$.
Analogous bounds   can be extrapolated to other 
scenarios (both strongly and weakly gravitating)
by way of the so-called covariant entropy bound \cite{bouyyy}.
So far, no violations of this bound are known 
given reasonable conditions on what constitutes physically allowable 
matter \cite{bouxxx}.

Although not  entropy bounds {\it per se},   dualities between
(bulk) gravitational and (boundary) field theories
provide another elegant realization of the holographic principle.
The most notable of these being   the duality that is known to exist 
between an  anti-de Sitter
spacetime and a conformal field theory of one dimension fewer; that is,
the celebrated ``AdS/CFT'' correspondence \cite{malxxx,witxxx,gubxxx}.
Such a duality is naturally holographic in the following sense:
If the correspondence is truly complete ({\it i.e.}, one to one), then
the lower-dimensional field theory
provides a strict means of limiting the amount of information that can be
 stored
in the bulk spacetime. Moreover, given that the field theory 
typically ``lives'' on a boundary of the spacetime, 
this entropic bound can
 be precisely related to  the area of some exterior surface.

An important distinction between the covariant  entropy bound
and these ``holographic dualities'' is that the former is believed
(at least hopefully)
to have universal validity  whereas the
latter  certainly does not.  That is to say, it 
is quite clear that a field theory dual does {\em not} exist for every
type of gravitational
theory. (Note, however, that this non-universality should not 
be viewed as a failure in the underlying principle. See
\cite{bouxxx} for further discussion on this point.)
One might then  be   inclined  to  wonder what fundamental principle
makes this  ``decision'' or, to rephrase, what exactly determines
 when  a duality does or does  not exist?  Although
the definitive answer will almost certainly require a
rigorous notion of  quantum gravity, it still behooves us
to see if progress can be made via semi-classical
considerations. In this regard, an important first  step  might  
be to establish  criteria for the existence/non-existence  of a 
holographic dual.

In a recent paper\cite{newcad}, Cadoni considered
various aspects of the  holographic principle in the context of
 two-dimensional dilaton--gravity theories.
(For other pertinent work, see \cite{Mig}. For a
 general review of two-dimensional gravity, see \cite{grumxxx}.) 
An interesting observation (actually, one of many) was that,
for a certain restricted  class of these dilatonic models, the 
black hole thermodynamic
behavior effectively mimics that of a free field theory.
More precisely, the  entropy ($S$) and
the energy  (or black hole mass, $M$) 
are related, for models  with a power-law (dilaton) potential,
in accordance with $S\sim M^{p/(p+1)}$ (where $p^{-1}$ is the
``power'').
It thus follows that, when  $p$ is a positive integer, this relation
mimics the thermodynamic behavior of a 
 field theory of dimensionality $p+1$.~\footnote{This notion of
black hole thermodynamics exhibiting an {\em  effective} 
dimensionality has been previously considered in the context
of critical-point behavior \cite{Cai}.} 

The main point of the current treatment is to demonstrate, quite rigorously, 
 that this
special class of two-dimensional theories can be identified with
a dimensionally  reduced form of (``strongly  curved'') anti-de Sitter gravity.
Moreover, the dimensionality of the anti-de Sitter spacetime,
$d=n+2$, fixes the parameter $p$ such that $p=n$.
Hence, one obtains an effective field theory of
dimensionality $n+1$, in compliance with the expectations
of the AdS/CFT correspondence.  Following the formal analysis,
we will explain how this outcome could be significant
in the context of our earlier discussion.

To begin here,
let us consider the gravitational action for an anti-de Sitter spacetime of
arbitrary dimensionality ($d=n+2 >3$). That is,~\footnote{The speed of
light, Boltzmann's  constant and Planck's constant will be set
to unity throughout.} 
\be
I^{(n+2)}={1\over 16\pi l^n}\int d^{n+2}x\sqrt{-g^{(n+2)}}\left[
R^{(n+2)} +{n(n+1)\over L^2}\right]\;,
\label{1}
\ee
where $l^n$ is the $n$+2-dimensional Newton constant  and 
$L$ is the anti-de Sitter curvature radius.
Note that 
$\Lambda=-{n(n+1)\over 2L^2}$ is the negative cosmological constant,
and so $L^{-2}$ measures the ``strength'' of the curvature.

Now, to reduce this {\em fundamental} action 
into that of an {\em effective} two-dimensional theory,
we  will utilize the following spherical {\it ansatz}:
\be
ds^2_{n+2}= ds_{2}^2(t,x)+ \phi^2(t,x) d\Omega_n^2 \;.
\label{8.5}
\ee
With this choice, it is straightforward (albeit tedious) to show that
equation (\ref{1}) transforms into 
%(also see, for
%example, \cite{newk,x5})
\be
I= {\V\over 16 \pi l^n}
\int d^2x \sqrt{-g}\phi^{n}
\left[R+ n(n-1)\left({(\nabla \phi)^2\over \phi^2}
+{1\over \phi^2}\right)+{n(n+1)\over L^2}\right] \;,
\label{9}
\ee
where  $\phi$, ``the dilaton'',  can be identified with the radius of the 
symmetric $n$-sphere, $\V$ is the dimensionless volume of a unit
$n$-sphere,  and all 
geometric quantities are now  defined with respect to the resultant
 1+1-manifold.

For convenience, let us (following \cite{newk,x5})  redefine
the  dilaton field as follows:
\be
{\psi}(x,t)\equiv \left[{\phi\over l}\right]^{n\over 2} \;.
\label{9.5}
\ee
The reduced action (\ref{9}) can then be re-expressed in the
following manner:
\be
I= {1\over 2G}
\int d^2x \sqrt{-g}
\left[D(\psi) R+ {1\over 2}\left(\nabla \psi\right)^2 
+{1\over l^2} V(\psi) \right],
\label{9.75}
\ee
where we have also defined
\be
{1\over 2G}\equiv {8(n-1)\V\over 16 \pi n} \;,
\label{9.80}
\ee
\be
D(\psi)\equiv {n\over 8(n-1)}\psi^2 \;,
\label{9.90}
\ee
\be
 V(\psi)\equiv
{n^2\over 8}\psi^{(2n-4)/ n} +{(n+1)n^2 l^2\over 8(n-1)L^2}\psi^2 \;.
\label{9.95}
\ee
Note that $G$ can be interpreted as the (dimensionless) gravitational
coupling  for the reduced theory.

Significantly,
the above  form of action  allows us to
 directly  implement  a special  field reparametrization 
that  eliminates the kinetic term \cite{kunxxx}.
More explicitly, following the  methodology of the cited paper,
we first redefine
 the dilaton, metric
and ``dilaton potential'' as follows:
\be
\Y=D(\psi)\equiv {n\over 8(n-1)}\psi^2 \;,
\label{9A}
\ee
\be
{\overline g}_{\mu\nu}\equiv \Omega^2(\psi) g_{\mu\nu}\;,
\label{9B}
\ee
\be
{\overline V}\left[\Y(\psi)\right] \equiv {V(\psi)\over \Omega^2(\psi)}\;,
\label{100}
\ee
where
\bea
\Omega^2(\psi) &\equiv& \exp\left[{1\over 2}\int 
{d\psi\over \left(dD/d\psi\right)}\right]
\nonumber \\
&=& {\cal C} \psi^{2(n-1)/n} =
{\cal C}\left[
{8(n-1)\Y\over n}\right]^{{n-1\over n}}\;.
\label{10}
\eea
In the last line [which, incidentally, made use of equation (\ref{9A})],
  ${\cal C}$ represents a  seemingly 
arbitrary   constant of 
integration. Nevertheless, this constant  can be fixed via physical
arguments (see \cite{newk}) so that  ${\cal C}= n^2/8(n-1)$.
For future reference, some straightforward evaluation
yields
\be
{\overline V}(\Y)=(n-1)\left[{8(n-1)\over n}\Y\right]^{-{1\over n}}
+(n+1){l^2\over L^2}\left[{8(n-1)\over n}\Y\right]^{+{1\over n}}\;.
\label{510}
\ee

In terms of the above parametrizations, the reduced action (\ref{9.75})  
now takes the simplified form
\be
I={1\over 2G}\int d^2x 
\sqrt{-{\overline g}}\left[{\overline \psi} R({\overline g})+ 
 {1\over l^2} {\overline V}({\overline \psi})\right ]\;;
\label{11}
\ee
and we will subsequently drop the cumbersome ``overline''
notation. It should be emphasized that this
form of the action is perfectly general.
That is, a generic two-dimensional dilaton--gravity  theory
can always be translated into the form of
 equation (\ref{11}) (given some modest constraints
on the initial action; see \cite{kunxxx} for details).

With the gauge choice $\psi = x/l\geq 0$,
the general  solution of this effective   action
reveals a static, Schwarzschild-like metric
\be
ds^2= - F(x)dt^2+ F^{-1}(x)
dx^2 \;,
\label{17}
\ee
where
\be
F(x)\equiv J(x)-2lGM\;,
\label{18}
\ee
with $M\geq 0$ representing the conserved mass  \cite{manxxx} and
\be
J[\psi(x)]\equiv \int^{\psi=x/l}V({\tilde\psi})d{\tilde\psi} \;.
\label{500}
\ee
(Note that the integration constant has already
been incorporated into  the observable $M$.)
Integrating  equation (\ref{510}), we specifically  obtain
\be
J(\psi)=n\left[{8(n-1)\over n}\right]^{-{1\over n}}\psi^{(n-1)/n}
+n{l^2\over L^2}\left[{8(n-1)\over n}\right]^{1\over n}
\psi^{(n+1)/ n}\;.
\label{520}
\ee

In a general sense, any form of $J(\psi)$  --- or alternatively
$V(\psi)$ ---
which admits black hole solutions (see \cite{cadxxx} for
the relevant criteria) leads to a well-defined  notion
of the associated thermodynamics (even though 
there is no strict analogy to the horizon area 
in a two-dimensional spacetime).
In particular, the temperature and entropy are
respectively identifiable as  \cite{lk3}
\be
T={1\over 4\pi l}V(\psi_h) \;,
\label{525}
\ee
\be
S={2\pi\over G}\psi_h \;.
\label{535}
\ee
Here,  $\psi=\psi_h$ locates the event horizon (assuming one
exists); that is, $F[\psi(x)]=0$ when $\psi=\psi_h$ or $J(\psi_h)= 2lGM$.

 Now, returning  to our  reduced anti-de Sitter model,
it is clear that an event  horizon does indeed exist;  inasmuch as 
$J(\psi)$ is strictly non-negative for any
admissible $\psi$.  It becomes a difficult
problem to solve analytically  for  $\psi_h$ under general circumstances.
Nevertheless, the situation noticeably improves
if one considers a   regime of ``small $L$'';
essentially, $L^2<<\phi^2$, where we recall that $\phi$
is the radius of the $n$-sphere.  [If we  consider the
relevant higher-dimensional solution ---  an
anti-de Sitter--Schwarzschild black hole in static coordinates --- then it is 
readily
confirmed that this condition translates into a lower bound on
the black hole mass: $M_{AdS-S} >> L^{n-1} l^{-n}$ 
(as also  discussed in \cite{newcad}.) 
Since our (implied)  semi-classical analysis
will, in all likelihood,  breakdown for sub-Planckian scales or
 when $L^2<l^2$, this bound
really indicates the  regime of a macroscopically large black hole
(in Planck units).  
Further note that, from the fundamental
or higher-dimensional
perspective,  this regime translates  into one
of high temperature~\footnote{This can  readily be observed  with
an inspection of the Hawking temperature  for an anti-de Sitter--Schwarzschild 
black hole. See, for instance, \cite{witty}.}
or, alternatively,
one in which the (negative)
curvature of the spacetime can not, even locally, 
be discounted.] In this case, one can verify --- with the help of
equations (\ref{9.5}) and (\ref{9A}) --- that the
right-most term dominates equation (\ref{520}), and so   
\be
J(\psi)\approx 
n{l^2\over L^2}\left[{8(n-1)\over n}\right]^{1\over n}
\psi^{(n+1)/ n}\; ; 
\label{530}
\ee
moreover,
\be
\psi_h\approx \left[n\over 8(n-1)\right]^{1\over n+1}\left[2 GL^2 M\over nl
\right]^{{n\over n+1}}\; ;
\label{540}
\ee
which is to say,
\be
S={2\pi\over G}\psi_h \sim  M^{n/ (n+1)}\;.
\label{550}
\ee

The interesting point about the above relation is
 that it effectively  mimics  an  $n+1$-dimensional field
theory with entropy $S$ and  energy $E=M$.  Given that our starting point is
$n+2$-dimensional anti-de Sitter space, this outcome can best be interpreted
as a  manifestation
of the AdS/CFT correspondence (although the {\em effective} field theory
described above is certainly not the one usually referred to
when this duality is discussed).
Nevertheless, the  relation  $S\sim M^{n/(n+1)}$ could have
similarly (and  rather more easily) been deduced by looking directly at the
small-$L$ limit of the (fundamental)
  anti-de Sitter--Schwarzschild metric \cite{newcad}.
Hence, it is reasonable to ask if
anything has really been gained by translating the theory into
a two-dimensional context.  
 
To address this  issue of relevance,
let us  first elaborate on a point made by
Cadoni \cite{newcad}.  Given a dimensionally
reduced action with the generic form  of equation (\ref{11}) [ignoring
the overlines],
one finds that, for a power-law potential or~\footnote{Note
that our   parameter $p$   is the {\em inverse} of the parameter $h$
used in \cite{newcad}.}
\be
V(\psi)\sim \psi^{1/p}\quad\quad {\rm with} \quad\quad p^{-1} >-1\;,
\label{560}
\ee
the entropy-mass relationship becomes
\be
S\sim M^{p/(p+1)}\;.
\label{570}
\ee
Hence, as pointed out in \cite{newcad},
one always obtains  field-theory-like thermodynamics  
whenever $p$ is equal to  a positive integer; with 
 $p+1$ being the dimensionality of this {\em effective}   
field theory.
The model considered  in the current letter can thus be viewed
as a concrete realization of this previously
observed phenomena.  More specifically, we have
found 
$V(\psi)\sim \psi^{1/n}$ (in the regime of interest),
and the effective field theory is consequently of dimensionality
$n+1$.  

In view of the above, we propose that
our analysis  provides  a clear illustration of
 how  dimensional-reduction
techniques may be employed  for 
identifying so-called holographic 
dualities (as discussed earlier in the letter).   
To elaborate, let us consider a hypothetical
(higher-dimensional) gravity theory  for
which an entropy-mass relation can
not be so readily  deduced from the fundamental metric.
The above techniques could  be purposefully utilized to
cast the theory into the reduced form of  equation (\ref{11}).
It then becomes a matter of inspecting the potential
for some viable regime (or regimes) in which 
  $V(\psi)\sim\psi^{1/p}$  such that $p$ is a positive integer. 
If such a regime does indeed exist,
 it would seem quite probable that some sort
of holographic duality will be in effect.
That is to say, two-dimensional dilatonic gravity 
may  yet provide us with  an intriguing means for clarifying
and, much more speculatively,  understanding  holographic
dualities.

Finally, it is interesting to note that, by way of the above reasoning,
 asymptotically flat space can {\em not} have any such field theory
dual. (Also discussed in  \cite{newcad}.)
To see this most definitively, consider the form of our
potential  [{\it cf},  equation (\ref{510})], but now
in the asymptotically flat limit of $L\rightarrow\infty$.
In this case, $V(\psi)\sim \psi^{-1/n} $ or $p=-n<0$,
and it becomes clear that a field-theory description is no longer
possible. Nonetheless, the inclusion of charge or
more exotic ``hair'' may possibly paint a different picture
(at least for some suitable choices of regime);
a topic which is currently under investigation.

\section*{Acknowledgments} 
\par
The author would like to thank M. Visser and D. Martin
for helpful conversations. The author also thanks
an anonymous referee for pointing out a tactical
error in the first version of the manuscript.
 Research is supported by
the Marsden Fund (c/o the New Zealand Royal Society) 
and by the University Research Fund (c/o Victoria University).
\par

\end{document}